# Title.

The need for an integrative thinking to fight against emerging infectious diseases

# Titre.

La nécessité d'une approche intégrative pour comprendre et lutter contre les maladies infectieuses émergentes

# Sub-title

Proceedings of the 5th seminar on emerging infectious diseases, March 22, 2016 - Current trends and proposals

# Sous-titre

Actes du 5e séminaire maladies infectieuses émergentes, 22 mars 2016 – Actualités et propositions

# Authors

C. Burdet[1,2], J.-F. Guégan[3], X. Duval[1,2], M. Le Tyrant[3], H. Bergeron[4], J.-C. Manuguerra[5], J. Raude[6], C. Leport[1,2], P. Zylberman[6]

(1) IAME, UMR 1137, INSERM, Université Paris Diderot, Sorbonne Paris Cité,
(2) Hôpital Bichat, Assistance publique – Hôpitaux de Paris, 75018 Paris, France
(3) UMR 5290 MIVEGEC IRD – CNRS – Université de Montpellier, Centre IRD de Montpellier, 34394 Montpellier cedex 05
(4) Science Po Paris, 75337 Paris Cedex 07
(5) Institut Pasteur, 75015 Paris, France
(6) Ecole des hautes études en santé publique, 35043 Rennes, France

# Corresponding authors

C. Leport

Adresse: UFR de médecine, site Bichat – 16, rue Henri Huchard – 75018 Paris

Tel: (+33) 1 57 27 78 68

Email: catherine.leport@univ-paris-diderot.fr

J.-F. Guégan

Adresse: Centre IRD de Montpellier, UMR MIVEGEC, 911, Avenue Agropolis – BP 64501 – 34394 Montpellier cedex 5

Tel: (+33) 4 67 41 62 05

Email: jean-francois.guegan@ird.fr

P. Zylberman

Adresse: 15, avenue du Professeur Léon-Bernard - CS 74312 - 35043 Rennes cedex

Tel: (+33) 1 45 27 04 88

Email: patrick.zylberman@gmail.com

## Running title

Integrative thinking on emerging infectious diseases

**Summary**

We present here the proceedings of the 5[th] seminar on emerging infectious diseases (EIDs), held in Paris on March 22[nd], 2016, with seven priority proposals that can be outlined as follows:

- Encourage research on the prediction, screening and early detection of new risks of infection
- Develop research and surveillance concerning transmission of pathogens between animals and humans, with their reinforcement in particular in intertropical areas (« hot-spots ») thanks to public support
- Pursue aid development and support in these areas of prevention and training for local health personnel, and to foster risk awareness in the population
- Ensure adapted patient care in order to promote adherence to treatment and to epidemic propagation reduction measures
- Develop greater sensitization and training among politicians and healthcare providers, in order to better prepare them to respond to new types of crises
- Modify the logic of governance, drawing from all available modes of communication and incorporating new information-sharing tools
- Develop economic research on the fight against EIDs, taking into account specific driving factors in order to create a balance between preventive and treatment approaches.

**Résumé**

Nous présentons ici les actes du 5[e] séminaire maladies infectieuses émergentes (MIE), qui s'est tenu le 22 mars 2016 à Paris. Les sept propositions prioritaires issues de ce séminaire sont les suivantes :

- Encourager la recherche sur la prévision, le dépistage et la détection précoces des nouveaux risques infectieux

- Développer la recherche et la surveillance sur la transmission des agents infectieux entre l'animal et l'humain, en les renforçant, en particulier dans les zones intertropicales, « points chauds » (« hotspots ») pour le risque infectieux, grâce à des aides publiques
- Poursuivre, dans ces zones, le développement et le soutien des actions de prévention et de formation des acteurs de santé locaux et favoriser la culture du risque dans les populations
- Prendre en charge les patients de façon adaptée afin de favoriser l'adhésion aux soins et aux mesures visant à réduire la diffusion d'une épidémie
- Développer la sensibilisation et l'entraînement des acteurs politiques et sanitaires, afin de mieux les préparer à la réponse aux crises de typologie nouvelle
- Réviser les logiques de gouvernance en s'appuyant sur tous les modes de communication disponibles, intégrant les nouveaux outils d'information collaborative
- Développer les recherches économiques sur la lutte contre les MIE, tenant compte des déterminants spécifiques afin d'équilibrer les démarches préventives et curatives

**Mots-clés**

Maladies infectieuses émergentes, écologie, microbiologie, une seule santé, crise sanitaire, stratégies et politiques d'aide au développement

# Manuscript

# 1. Introduction

The objectives of the 5th edition of the Val de Grace School's seminar were to take a global, integrative approach to the emergence of new infectious disease agents, putting these in perspective relative to other types of risk, as well as studying crisis situations and the breaches brought on by the occurrence of emerging infectious diseases (EIDs). Two key conferences respectively opened and closed this annual seminar; the first was given by Dr. Peter Daszak (President of the EcoHealth Alliance and program director of USAID-EPT-PREDICT), the second by Dr. Patrick Lagadec (former research professor at the Ecole Polytechnique, Palaiseau).

Planned for an anticipated up to 200 participants, this seminar is designed for decision-makers, experts, medical doctors and scientists interested in human and animal health, social sciences, environmental sciences, prospective analysis, biosecurity and defense.

This day of presentations and debates is presented under the auspices of the French Social Affairs and Health Ministries, as well as that of the Environment, Energy and Marine Affairs. It is organized in the framework of a multiple partnership with ANSES (the French Agency for Food, Environmental and Occupational Health & Safety), the Health Chair of Sciences-Po, the EVDG (Val de Grace School) from the Armed Services Health Division (SSA), the School for Advanced Studies in Public Health (EHESP), the High Council for Public Health (HCSP), the Pasteur Institute of Paris, the French National Research Institute for Sustainable Development (IRD), the Thematic Multi-organisms and Public Health Institutes (ITMOs) - Immunology, Inflammation, Infectiology and Microbiology Institute (I3M) and Public Health Institute (ISP) from the National Alliance for Life Sciences and Health (AVIESAN), the French Microbiology Society (SFM), the French Infectious Diseases Society (SPILF), the Paris - Diderot University and the BioMérieux Foundation.

## 2. Current issues: presentations and debates

2.1 Keynote conference - Understanding the ecology and economy of pandemics

*Speaker: Peter Daszak (The EcoHealth Alliance)*

*Moderator: Jean-François Guégan (IRD)*

Peter Daszak remarks today can be summed up in two primary messages, both personal and that of scientists from the EcoHealth Alliance. First, we must increase our research capabilities in order to understand the direct and less-direct causes of emerging infections if we hope to fight them once they become responsible for epidemics or pandemics. Secondly, this presentation was illustrated with a few projects concerning the economics of EIDs which reveal the extraordinary costs they incur for national economies. There is at least one obvious reason to speak of this, which is that politicians and public decision-makers are in fact quite sensitized when the economic damage engendered by the latest epidemics and pandemics are explained to them. If one takes, for example, the SARS-Cov pandemic, it led to a 1 to 2% decrease in gross domestic product in several Southeast Asian countries, for an estimated overall cost of 30 to 50 billion US dollars, relative to the total number of approximately 800 people worldwide who were affected. The appearance of new EIDs appears both more frequent and also vaster in terms of number of people affected these last years, many of these diseases appearing in developing or low-income countries. Taking the case of the Ebola virus (EBOV) epidemic which broke out in West Africa in 2014, it was much more widespread than any other epidemic to ever occur in Central Africa (400 people affected during an epidemic). It is very difficult in the United States – where Peter Daszak is working – to get public decision-makers interested, as is true with the population which feels very distant from such problems. EcoHealth Alliance has tried to draw the media and the public's attention to it, in vain! They only received one single email from the management of the Boston airport, informing us that our work on the risk of Ebola virus propagation via transcontinental air transport was unfounded and that the Boston airport couldn't experience this type of threat. About a month later, when an American citizen was repatriated for treatment, the public began to panic; the media, particularly the TV channel CNN, exaggerated in broadcasting the issue, and the government took several decisions, notably regarding assistance and monitoring of international airports. These measures were decided upon no scientific basis in a moment of panic, and it seems that with every new worrisome EID, it's the same story! The measures taken are often disproportionate to the seriousness of the phenomenon. Human demography has exploded in recent years, and populations today are concentrated in megalopolises. This tendency is even more marked in tropical areas in the South where there is also an important biodiversity of animal species. Therefore, there exist today more opportunities for a virus or bacteria to pass from an animal to a human, then to propagate among the human population. International air transport makes possible the spread of these new infectious risks on a vast scale. What are governments doing faced with this type of threat when the demands on the part of the population are ever-increasing? Vaccines are also considered by the former as a weapon of total destruction, and by the public as a widely-available miracle tool. Neither is truly aware that it takes an average of 10 years to produce a vaccine. Also is it the right strategy in that it banks more on cure than on anticipation? In the U.S., President Obama opted for a development aid policy through USAID favoring training and improvement of human and technical capacities to prevent future EIDs in the most needy countries. The American Congress, however, did not follow this path, instead directing funds allocated to USAID towards research on a vaccine for and on epidemic management of the Ebola virus crisis.

Recently, in collaboration with economists, EcoHealth Alliance modeled the risk of a new emergence in a situation identical to that of Ebola in 2014-15 in West Africa, estimating the economic damage caused, and attempted to deduce the economic cost of a health policy based on epidemic management as opposed to one favoring prevention and training. According to their estimates, a budget of 5 billion US dollars seems to suffice to prevent a new epidemic in West Africa in opting for the second solution. Their simulations obviously include many underlying hypotheses. All the same, this analysis indicates that if we were to quickly make available 1 billion of the 5 billion dollars for the purchase of equipment, the setting up of working laboratories on site, field hospitals, sending doctors and nurses to the area, training of our partners in affected countries - as military medical services know how to do - such a strategy would represent a real capital investment in managing epidemic propagation. Unfortunately this was not the option chosen by U.S. current governments, which we can only regret.

Now let's discuss the ecology of Ebola virus transmission. The bat species *Pteropus* are possibly the host reservoirs of Ebola virus. As for the Marburg fever virus we are now certain that these bats are in fact the reservoir. The Ebola virus is clearly present in a few giant bat species in Africa. It can also circulate among primate species or indirectly through other infected mammals. We also know which species of bats are affected or not. Even if the reservoir question is still under debate, we know a great deal about the rainforest circulation of this virus and the circumstances of its spreading in the human population with high-risk groups such as hunters. Based on this, is it possible to prevent the virus' transmission to humans? Although it poses a complex question it is nonetheless possible to offer some avenues of reflection. William Karesh, vice-president of EcoHealth Alliance and instigator of the OneHealth concept, carried out a research project in the Democratic Republic of Congo, which consisted in educating villagers, and particularly hunters, regarding the danger of recovering primate corpses from the forest. This program was a real success, as the villagers changed their behavior and the eating of monkey meat. This represents a low-cost prevention approach, which in the end functions well if properly implemented. The only real solution when faced with this type of health threat is to work in conjunction with populations, and to treat questions of poverty, equality, and use and transformation of land together. These are arduous and complex tasks which must be carried out over the long term. This type of message is extremely difficult to make heard among politicians and the public, and obviously more complex that announcing or clamoring for a vaccine.

We are faced with primordial questions which are more or less difficult to answer. Are we witnessing an increased frequency of EIDs appearance? Are there more cases today? Can we predict patterns, or rules, of emergence? Are there geographic areas in which these emerging infections are more frequently or widely observed today? Do we possess the scientific ability to predict exactly where the next infectious epidemic will start? Evidently, if we were so able we might allocate the financial resources and technical and scientific support to suspected areas. At the moment we do just the opposite, or we do nothing, and we are disconcerted at each new emergence. In order to convince governing bodies and public decision-makers, we must demonstrate to them that predictive approaches are less costly than curative ones. Using approaches inherited from ecology and biogeography EcoHealth Alliance has adapted the same principles to understanding the ecology and spatial distribution of EID principles observed over the last decades. Their work shows that of the nearly 400 new EIDs which have appeared, the so-called zoonotic ones originating in wild animals have been in net increase in the last 40 years; 3 out of 5 new infections which appear each year originate in wild fauna. What's more, the risk of infectious transmission is highly and doubly correlated with human density, as are these diseases of animal origin to the areas' biodiversity. They called these areas "hot-spots" for disease risks, mostly situated in intertropical regions. They are advocating for these specific zones to be those on which they concentrate our research efforts as well as international public aid for development. In fact, it is not only biological diversity in animals which must be taken into account, but also the evolution of natural ecosystems and their disappearance over recent years. Deforestation and land use changes by populations are driving forces in the appearance in new EIDs. Therefore, international politicians must better link economic strategies to those of habitat and biodiversity coordination if we wish to avoid new pandemics.

In further use of the formalism of economic models, they have shown that a stable political strategy appears once joint simulations were conducted on economic damage due to a pandemic health crisis and preventive decision-making. Even though the financial cost of preventive strategies may initially appear great, an optimal solution based on prevention shows itself over the mid- and long-term to be finally less costly than cure and control-focused strategies. We have need on an international level for the equivalent of the Intergovernmental Panel on Climate Change (IPCC) which, in the same way as does the IPCC on climate change scenarios and their impact, would concentrate on EIDs and their sociopolitical, economic and environmental consequences. The American development aid agency USAID did not originally give priority to EIDs, but the different public health crises brought on by the H5N1 bird flu virus in recent years have led USAID to modify their strategic orientation, notably with regard to the economic weight they bring to bear on regional economies. The fact that these epidemics appear first in countries situated in strongly species-rich intertropical areas which are developing or low-income, requires a reconsideration of our Western policies on development aid in order to better include these notions. For ten years USAID funded a program on new emerging infection threats, for a total of 1.3 billion U.S dollars. At EcoHealth Alliance we have collaborated with this initiative through participation in the PREDICT research project with funding of 45 to 50 million dollars. The goal of the PREDICT program was to identify microorganisms potentially pathogenic for human populations, and which are hosted in animal reservoirs. We took particular interest in three animal groups, not only because it was impossible to work on all the animal groups present, but also because the three groups we chose - primates, rodents and bats - were recognized as being major reservoirs of agents which are pathogenic for humans. These three groups alone make up 75% of the world's mammal species. Within the PREDICT program we applied our knowledge of the high-risk emergence zones to draw samples from a large cohort of these three animal groups. Over the first five years of the program we sampled 56,000 animals, trained 2,500 scientists and medical and administrative personnel, and discovered more than 1,000 new viruses belonging to families of viruses known not to be infectious to humans. This in itself represents a major finding! Obviously the discovery of a microorganism does not in and of itself indicates that a new EID might appear. Their work in Mexico on bat species demonstrated the existence of a dozen viruses which are very like the MERS-CoV responsible for the respiratory syndrome in the Middle East. They therefore believe that the MERS-CoV is not hosted by dromedaries but rather by bats, as they found it in our work in Mexico. Some of these new Coronaviruses may potentially be pathological for humans. It is clearly impossible to identify all viruses present on the planet. It would also be necessary to do the same for bacteria, parasitic fungi, and protozoa. As they are innumerable, it is preferable to make strategic choices regarding microorganism research and that on the highest-risk animal groups. Based on our knowledge of currently known viruses in Bangladesh bats, they extrapolated this data using species curve rarefaction, and capture-recapture techniques well known to ecologists, to estimate the total number of viruses to be expected in the totality of known mammal species in the world. They reached a value of 320,000 new viruses which remain to be discovered. Were we to carry out biodiscovery research on these viruses, one could catalogue them, classify them in relation to already-known viruses, in particular those known to be pathogenic to humans, and through comparative genome studies assess their pathogenic potential. We are currently in the second phase of the US PREDICT program, with an important accent on our manner of working. In fact we are currently concentrating more on factors involved in emergence and seek particularly to understand three such factors: habitat change and use, intensive agriculture, and the commerce of biodiversity. US development aid policy also focuses on three strategic areas: North Africa, West Africa, and continued research activity in Saudi Arabia. These geographic choices are shaped by recent events around the MERS-CoV and Ebola viruses. Especially in the case of MERS-CoV, dromedaries are certainly involved in the transmission cycle of the virus, but we believe that its true reservoir is the bat. Thus using the many data at our disposal we were able to show that the infectious risk of MERS-CoV for humans is not great in Saudi Arabia, but it is in other areas of contact between bat, dromedary and human populations, particularly in the horn of Africa, especially in Somalia. Somalia is currently a politically disjointed country, where public health surveillance and care are greatly lacking or simply non-existent. For example it is currently impossible to state how many MERS-CoV cases there are in Somalia, or how many might exist. Our scientific objectives are therefore to better understand the behaviors at the junction of wild or domestic fauna and human populations.

For example, understanding human behaviors and practices at the interface of tropical forests and villages could help us to interpret how zoonotic transmission happens. Through acting upon these behaviors and habits we could lower this type of emerging risk. Currently EcoHealth Alliance has research sites in Uganda, Malaysia and Brazil; they are investigating the role played by habitat changes such as deforestation, human interaction with biodiversity through behavior and use. In Manaus in Brazil for example, the maximum risk of new infection is not in the heart of the city where the major markets are located, but in the peri-urban areas where agriculture and ranching are developing. These zones which ecologists call « ecotones », or transition zones, are located near strongly species-rich regions, with high concentrations of livestock which can come into contact with wild fauna, and which are furthermore inhabited by ranchers and farmers. These regions which are now located everywhere in the intertropical world for the purpose of feeding urban populations, are generally those where new infections appear, and where future pandemics will also most likely appear. At the heart of this research is the priority of understanding behaviors, habits and practices of populations with the goal of changing these factors. This is a long-term endeavor which requires developing approaches in the field for communicating with local populations, and for developing community participation in our own research, which we also consider an excellent means of education.

## 2.2 (Re)-Thinking public health risks: the lessons to be learned from sociological studies on technological risks about infectious risks

*Moderators: Henri Bergeron (Chaire Santé de Sciences Po), Jocelyn Raude (EHESP, IRD)*

### 2.2.1 Risks and the limits of the State

*Presenter: Olivier Borraz (CNRS, Sciences Po)*

O. Borraz recalled the fact that we live in a society of risk. It is not that the dangers surrounding us are more numerous or more fearsome than before, but simply that the notion of risk plays a central role in public policies, in public and private organizational management, and in the controversies around new technologies. Genetically modified organisms, mobile phones, nuclear waste, urban sanitation sludge; the activities now considered health or environmental risks are countless. This categorization puts public authorities in a position of having to ensure the safety of populations, even as the State itself sometimes represents a risk factor. It is therefore essential to understand how an activity becomes a risk, and how it is then managed by public authorities as well as by companies, associations and local conglomerates. Risk from its identification to its management, from its highlighting to its instrumentalization becomes a tool linked to the emergence and expansion of a welfare state. It is used by politicians to justify on the one hand its lack of involvement in risk management, on the other the seizing of power over sectors from which they had previously disengaged. This political power seizure plays on politicians' selective dissemination within society of risk identification and the knowledge required for its management.

Two different processes exist by which risk represents an organizing principle for political power, a means to assist and contribute to the definition of the State's limits: « putting at risk » and « regulation through risk ».

« Putting at risk » refers to all processes by which an event is described as or constitutes a real or predicted danger, and which thus is categorized as a risk [1]. There are many events, objects, and situations which have historically been categorized as « putting at risk » and which can be studied by sociologists (illness, divorce, food crises, unemployment, nuclear risk, chemical substances, technological risks …). This putting at risk can be a product of the State or of interest groups, under the influence of lobbying. This has been part of the organization of modern societies since the 20th century, and in particular following World War II with the expansion of the welfare state which took advantage of different factors putting the population at risk in order to enlarge its sphere of activity and thus its power in the name of a protective mission. This welfare state built itself around risk management in creating and organizing agencies, action plans, drills, and monitoring mechanisms for their application through nationwide inspection bodies.

The promotion of « risk instruments » and a « risk-based regulation » approach makes it possible to deal with risk-creating events. For the last two decades, in a context of decreasing means and in reaction to public-health crises which have led, according to some, to an overprotective state, a new tendency has emerged of governments using « risk instruments » to better allocate means and to decrease the State's hold. Despite their differences these two approaches are to the same end, which is to define - or perpetually re-define - the limits of the State. « Risk instruments » contribute in each « putting at risk » situation to determining the State's involvement, the reason for its having competencies and resources, and therefore also the State's limits as opposed to that which concerns the private sector and individuals. The outcome of this defining as « putting at risk » and of the application of these risk instruments may be used by the State either to invest, or to disinvest; in the latter case for example entrusting risk management which formerly would have been the realm of institutions to local or private agents or to individuals.

In the same way, risk instruments (or risk-based regulation or using risks to rationalize public intervention) may lead to State disinvestment or, on the contrary, be used by the State to seize back the reins in certain areas. Paradoxically risk instruments serve in this instance to re-centralize or to « remote-control » areas to which the State had granted greater autonomy (universities or hospitals, for example).

Thus the way in which Western societies manage risk, and the way in which these risks - their identification, how politicians and individuals perceive them, and their management - transform the links between the State (the political powers-that-be) and civil society are important to understand for those faced with risk management, especially in public-health risk management. What consequences these risk technologies have from one country to another thus greatly depends on institutions, State structure, professional organizations and their interrelationships, the balance of power and the forces involved, and the legal structure - which may be more or less open to interpretation according to how it is drafted (more or less restrictive laws). Risk management is, after all, at the core of the State and has always been central to its transformations.

## 2.2.2 How climate change becomes conscience in the crucible of complexity

*Presenter: Alfredo Pena-Vega (EHESS, CNRS)*

Climate change can be interpreted, according to A. Pena-Vega, as a complex system such as Edgar Morin defined it in his « Introduction to Complex Thought » [2]. The stakes involved in it are multidimensional, and suppose at once the ideas of transformation, uncertainty and unexpectedness, unpredictability and crisis …

It henceforth seems necessary to arm ourselves with instruments enabling us to understand this reality. The common denominator of such instruments is rationality. Because climate change involves stakes which are climatological, economic, social and health-related, as well as questions of governance, ethics, biodiversity etc. Cross-disciplinarity appears to be one form of rationality which allows us to avoid the dangers of « fear-mongering ». A study was conducted among 12,000 high school students from 20 different countries, of which the objective was to understand how younger generations see climate change. The results are quite varied. If a minority of responders question the existence of climate change - with an argument largely based on a social construct communicated by the media - 90% of responders claimed to be concerned and aware of the negative consequences of climate change. The vast majority consider it to be a threat to the survival of humanity, in particular due to the multiplication of natural disasters, dwindling biodiversity, the spread of new infectious diseases, notably those with vector-borne transmission, and the inequalities it fosters between humans.

## 2.3 Reflecting on and steering « out of framework » public health crises in a pulverulent universe. Presentation and exercise

*Speaker: Patrick Lagadec (International Consultant)*

*Moderator: Catherine Leport (Université Paris Diderot - INSERM)*

We are currently in the « age of the unthinkable ». Today's world constantly exposes us to new crisis situations which we must learn to confront. These situations are all the more difficult to manage in that they most often occur « out of framework», or within a framework in which it is difficult to define the outlines delimiting an increasingly « volatile » environment. Whereas our benchmarks in terms of crisis management are structured according to typologies (natural/biological/social disaster), the boundaries of current crises are unclear, and their typologies intertwined. Unexpected and unpredictable, they emerge within a context where uncertainty reigns and where the response is organized according to the logic of competitive leadership. Information no longer follows a downward flow from the State to the public, but circulates in a collaborative manner via widespread connectivity, social networks, which compete with institutional media which are sometimes outdated. It is therefore difficult for decision-makers to circumscribe the areas of operation, to isolate causes, and to distinguish the components which permit specific, technical and successive interventions. Good judgment thus becomes key.

Public health crises, and especially those concerning EIDs, have pointed up the difficulties in predicting their appearance and development. Comparisons and connections between these and other types of crises (natural, technological, industrial …) and how they are interpreted can thus prove useful in public health crisis management.

The unattainable goal of predicting such crises makes it necessary to prepare to meet with unexpected events during the ongoing management of the crisis; it is no longer a question of predicting the unpredictable, but preparing to deal with it. In the context of our intricate, complex society, coordination and communication are of course necessary, but it has also become imperative to have a thorough knowledge of the steering process.

In « out of framework» situations, strategic thinking capability takes precedence over the quality of available technical expertise. This can take the form of an « express think-tank », a group made up of diverse members capable of and trained to work together in situations of uncertainty. Whereas in a « classic » situation the command functions is a pyramid, in these « out of framework» situations, collaborative and flowing cooperation is required, without leadership's impeding the efficaciousness of the response. In the case of Hurricane Sandy in the U.S., in 2003, several work teams were set up in order to handle the unprecedented nature of the situation: « real-time innovation », « immediate flaw detection », « emergency support functions ». These parallel work teams made it possible to optimize management of the disaster by limiting disastrous consequences. It seems essential to develop such network-based crisis management in France.

2.4 Interactive session - conditions and factors of emerging infectious agents in humans.

*Speakers: Maria van Kerkhove (Pasteur Institute, Paris), Allison McGeer (Mount Sinai Hospital, Canada), Benoît Guéry (CHRU, Lille)*

*Moderators: Jean-Claude Manuguerra (Pasteur Institute, Paris), Patrick Zylberman (EHESP)*

P. Zylberman recalled to mind the definition of an EID as an unexpected infectious - or presumably infectious - phenomenon, affecting humans, animals or both. According to the definition used in the High Council of Public Health's 2011 report [3], this can entail an infectious clinical entity which has just appeared (« true » emergence), one previously identified (known emergence) or a known infectious disease whose incidence has increased or whose characteristics have changed (re-emergence). The HIV epidemic in the 20th century or the SARS-CoV epidemic in the early 21st century are examples of true emergence. The emergence of Hepatitis C corresponds to a known clinical entity whose etiological agent was identified at the end of the 1980s. The measles and West Nile virus outbreaks on the American continent, in the 19th and end of the 20th centuries respectively, are examples of re-emergence.

Nathan D. Wolfe has defined 5 stages of transformation of an animal pathogen into a specifically human pathogen, resulting in a « true » emergence, with the possibility of an evolutionary interruption at each stage. Stage 1 corresponds to a situation in which a known virus in animals has never yet been detected in humans in natural conditions. In Stage 2, the virus known in animals is capable of infecting humans in natural conditions, but without the capability of person-to-person transmission. In Stage 3, some cycles of secondary person-to-person transmission are possible. In Stage 4, the virus circulates among humans through several secondary person-to-person transmission of varying duration. Stage 5 is reached when the virus becomes exclusively human, and also contagious.

On a population-wide scale an epidemic goes through four phases: introduction, propagation, amplification, and regression of the infectious phenomenon. The propagation phase is that during which there are the most widespread and frequent sites of infection. It corresponds (particularly for viruses) to the adaptation of the pathogen to its new host with person-to-person transmission taking effect little by little. It is often at this stage that the epidemic phenomenon is detected, sometimes with a considerable delay relative to the introduction of the pathogen. Propagation can take place not only by contiguity but also across vast distances. Humans thus play a role through their activities, and are what S. Morse refers to as «microbial traffic engineers» [4].

What are the major factors in the emergence of new infectious pathogens in humans? A great deal of progress has been made in improving the rapidity in identifying new viruses, as J.-C. Manuguerra points out. The time delay between the individualization of a new infectious nosological entity and the identification of the causal agent thus continues to shorten, from 14 years for hepatitis C in the 1970s to 2 years for HIV in 1983, and to 6 weeks for SARS in 2003; even less for MERS-CoV. Even so, the discovery or the knowledge of the pathogen's existence does not provide all the answers about the risks posed for or by a host species. Various behaviors of pathogens can nonetheless be observed, in particular the pathogen's adaptation may require passing through several intermediary host species before adapting to its final host. Moreover, the epidemic potential of a discovered virus is difficult to determine. Among the very numerous known arboviruses, most are of anecdotal importance for human pathology, and little has been undertaken upon their discovery in preparation in case the pathogen were to become epidemic. This is true in the case of the Zika virus, first isolated over 50 years ago, and currently responsible for a major epidemic in Latin America and in the Caribbean.

The question of the time lapse between the beginning of an epidemic and identification of the pathogen is progressively slipping toward that between the beginning of the phenomenon and its detection by the health system. This period appears to be critical for eradicating the development of an epidemic. Is it possible to narrow the time gap between the beginning of an epidemic and its detection by the healthcare services?

Early identification of an epidemic-level incident mobilizes all health services professionals, both in regards to human and to animal health, according to M. Van Kerkhove. Human and animal health and the state of ecosystems are inextricably linked, and it is believed that nearly 60% of EIDs, including those re-emergent, are of zoonotic origin. It is through this exploration of the connections between animals and humans that the means of transmission and propagation of MERS-CoV within the human species could be revealed. This virus is an example of a potentially epidemic emergence.

Schematically, during the MERS-CoV epidemic, a limited number of person-to-person transmissions, and sporadic cases between dromedaries and humans, were observed, with a certain degree of increase seen in healthcare facilities, sometimes considerable, as was the case recently in Korea. During the epidemic 75% of the MERS-CoV cases were reported in Saudi Arabia, and of 1600 cases reported to the WHO task force, 60% were considered to be primary cases, that is, contracted from an animal host source, and 40% to be secondary cases, or acquired through another human case. Every primary case represents an opportunity to understand how the infection was contracted. It gradually seemed necessary to launch a veterinary investigation as soon as a case was diagnosed. This led to the development of animal surveillance, which made it possible to reveal the seropositivity of certain dromedaries, as well as active excretion of the virus in their environment, which made possible its transmission to humans. This improvement in detection of emerging pathogens and early epidemic detection calls for the improvement as well of the veterinary surveillance network, the so-called OneHealth approach. In an area with limited resources, where all animals cannot be tested, it is imperative to concentrate on those areas with high concentrations of animals, such as slaughterhouses, and on areas in which humans come into close contact with animals and thus represent a high risk of transmission (see conference by P. Daszak). The data collected in these areas of frequent contact between humans and animals can be utilized to issue recommendations for at-risk populations in order to reduce transmission. Once the virus has acquired person-to-person transmission capability, its control is far more complex due to the rapidity of propagation following its introduction into the population. It therefore becomes difficult for the healthcare system to improve detection of the epidemic. Establishing health policies to optimize the case reporting system is thus critical.

How to provide care to patients in the case of new EIDs? In an epidemic situation, the treatment of affected patients usually is secondary to the need to control the disease, as A. McGeer stated. Nevertheless, an adapted patient care procedure can change the evolution of an epidemic, to varying degrees according to the pre-existing healthcare infrastructure in the given country.

In North America, where the services for the monitoring and managing of infectious diseases and public health are separated from the healthcare system, the organization of care for infected patients is usually left to the physicians. The healthcare system is in fact organized around individual patient care, and not oriented toward an approach of global individual and collective care providing. In the case of EIDs, care of affected patients becomes a political issue in a country graced with a public health system. Health is seen as a human right, and governments are judged not only by their ability to prevent and manage epidemics, but also according to their management of care provided to ill patients. The role of the public healthcare system is therefore to advise physicians and to develop recommendations for the detection of cases and their homogeneous treatment. These guiding principles make the physicians the kingpins between healthcare structures and the treatment of patients, and the public healthcare system. Better patient care provision thus improves epidemic response. In fact, treating ill subjects also allows the risk of person-to-person transmission and propagation to be reduced.

This treatment role attributed to physicians can be variously interpreted in countries in which the healthcare and public health system are underdeveloped or lacking. In fact, the arrival of healthcare personnel and the setting up of precautionary measures required to contain the epidemic may be experienced as an intrusion. The lack of comprehension and communication between medical staff and the affected community, as well as the potential lack of comprehension of the measures set in place can prompt affected individuals to hide, due to the uncertainty of their fate. This was the case for example in West Africa during the Ebola virus epidemic, during which affected patients were sometimes hospitalized far from their villages without care being taken to inform their families of their clinical progress and outcome. This led to sometimes-violent rejection of the healthcare personnel, which interfered with the measures meant to control the epidemic.

Another element to take into consideration in the care of affected patients is protection of the healthcare personnel, which has become a recurring problem. Most of the microbial forms with epidemic potential propagate within the community. For a long time hospitals were not troubled over the risk of nosocomial transmission of EID agents. Care providers were not aware at the time of the risk of contagion that they themselves ran when providing care to patients. This awareness happened during the recent SARS,

MERS-CoV and Ebola epidemics, which shared certain physio-pathological characteristics different from those involved in previous epidemics. In fact the peak of virus excretion for measles, chicken pox and influenza generally occur before or upon the appearance of symptoms, and the risk of transmission is virtually nil when the patient arrives at hospital. SARS, MERS-CoV and Ebola virus have a different viral excretion rhythm: symptoms appear with a weak viral load, and their intensity increases with the level of viral excretion, which reaches its maximum just when the patient requires the most care. The hospital thus becomes a center for the propagation of the pathogen. The hospital system is therefore in danger of breakdown, as it is both the pole for patient care and the new center of the infection's transmission.

These new pathogens therefore require a re-thinking of hospital design, in order to optimize both control of epidemics and patient care.

What has been, or is, the extent of nosocomial infection's role in the transmission of SARS-CoV and MERS-CoV? The new pathogens require us to re-evaluate modes of prevention for nosocomial transmission, confirmed B. Guéry. The 2003 SARS-CoV epidemic serves as a good example of what was learned about the intra-hospital transmission of these new pathogens. Person-to-person transmission of SARS occurs through droplets, physical contacts and airborne pathways. The SARS-CoV transmission rate to healthcare providers exclusive of invasive procedures has been estimated at 21%. The main risk factor identified was the lack of protection of the provider's airway through wearing a mask, with an odds-ratio estimated at 13 (3-60). Wearing scrubs and handwashing were also associated with lower transmission risk.

SARS is a disease which appeared in 2003 in Guangdong Province in China, then in Hong Kong, where numerous primary and secondary cases occurred. In total, according to the case index, 716 secondary and tertiary cases occurred, of which 52.3% among healthcare providers. Beyond standard hygiene measures, studies conducted on affected healthcare providers revealed that certain categories of personnel, such as technicians and nurse's aids, show an infection rate twice that of the nursing staff, and 6 times greater than the medical personnel. These studies made it possible to identify infection risk factors generally not taken into account in the fight against nosocomial transmission of pathogens: a significantly, greatly increased risk (odds-ratio 7.3) was noted in care providers having used precautionary measures against SARS transmission for less than 2 hours, as was the case with those not having understood the protective measures (odds-ratio 3.1).

The factors identified as influencing transmission are patient viral load and patient index distance. The ideal conditions for transmission to occur are those of an infected patient excreting large quantities of virus, presenting with a certain number of co-morbidities capable of masking the initial profile, and the existence of multiple close contacts with high-risk procedures such as oro-tracheal intubation, performing a fibroscopy, or the administration of treatments through nebulizers. The idea of a « super-excreter » patient was also identified during recent respiratory virus epidemics. This concept could play a crucial role. Usually it concerns cases of very serious infection occurring in patients with several co-morbidities.

In Beijing in 2003, the SARS case index was also associated with a large number of secondary cases (76 cases, of which 12 among healthcare personnel). As with MERS-CoV, despite a relatively low basal reproduction rate ($R_0$), a large number of care providers were infected during the epidemic. This is what happened, for example, in Abu Dhabi, where 65 cases of MERS-CoV were diagnosed, of which 42% among healthcare providers. Each case of provider infection was followed up through an epidemiological study, and each time an obstacle to the isolation of the patient and to the application of hygiene practices was noted. Over 80% of the MERS-CoV cases identified in Korea were thus traceable to 5 « super-excreter » patients. This notion remains questionable, as it is reductionist and could lead to the identification only of patients in this category, to the neglect of transmission risks associated with other patients. It is probably more fair to speak of « hyper-excretion events » which implies that each patient is at maximum risk, and should be treated using precautionary measures.

In conclusion, the intra-hospital control and transmission of EIDs can only occur in connection with the development of precautionary standards, which should be ongoing over time, and should be applied by all healthcare providers. It is imperative to ensure that caregivers are adequately and regularly trained, and that they constantly keep in mind the importance of isolation of all infected patients. To achieve this, it is probably necessary to resort to specialized units, in reference hospitals, in conjunction with clear decisions at the national level.

## 3. Synthesis and proposals

<u>Axis 1: Global health and ecological approach</u>

An integrated approach to health in face of the globalization of risk has been developing over the last few years. The « OneHealth/EcoHealth » concept, or global health, takes into account the fact that human health, animal health and environmental health are inextricably linked, especially in regards to EIDs, exposure to which is fostered by the multiplication of transcontinental travel, many instances of human-animal contact, and intensive farming and ranching. Many recent examples have made it possible to establish the key role played by animal biodiversity in the introduction and transmission of pathogens within human populations. Whether it be the role of bats in the 2014 Ebola virus epidemic, or dromedaries in the 2012 MERS-CoV epidemic, the crucial role of animals and of human-animal contact - being wild or domestical animals - in triggering an epidemic has recently been emphasized. Primates, rodents and bats are the three mammal groups most likely to be at the origin of future pandemics due to the high proportion of viruses which they share with humans. The inclusion of fields which appear quite unrelated (infectious diseases, animal health and ecological and environmental sciences) should thus be pursued and improved.

It has been possible to establish models which allow the prediction of emergence tendencies in infectious diseases, and certain geographical areas at high risk for emergence have been identified: Central Africa and West Africa, Southeast Asia, Central America. These areas correspond to those at high risk of propagation due to underdeveloped or deficient public health systems, and to the absence of epidemiological surveillance. Tools necessary for effective prevention of future epidemics are now available. These are all the more critical in that recent increasing tendencies raise fears of a multiplication of the number of emergent epidemics in future. Beyond the fight against pathogen propagation, its introduction into the human population is in fact a key step against which « battle plans » can be drawn up.

In order to perfect pandemic response, it is necessary to improve the coordination and interconnection between individual and institutional participants, such as healthcare providers and public health systems. A global response approach (OneResponse) should be reflected upon and developed. On an institutional level, a first step in bringing together the fields of environment and animal health occurred in 2010 in France with the creation of a national French Agency for Food, Environmental and Occupational Health & Safety (ANSES), originating in the French Agency for Food Safety (AFSSA, which also includes the National Agency for Veterinary Drugs) and the French Agency for Environmental and Occupational Health & Safety (AFSSET). Another case of bringing together the areas of surveillance, prevention and human health intervention occurred in 2016 with the creation of the French Public Health Agency, with the merger of the Health Surveillance Institute (InVS), the National Institute for Prevention and Health Education (INPES), and the Organization for Preparedness and Response to Health Emergencies (EPRUS). The relations between these two new institutions should be developed in order to provide a better-coordinated response to future health crises.

Axis 2: Preventing and anticipating epidemic propagation

The response to an EID should take into account not only factors linked to EIDs, but also to a constellation of political, economic and socio-cultural constraints. The decision to put in place such a battle is in fact a political decision which involves, beyond the scientific aspects, the intervention's impact upon the popularity of the acting political powers-that-be. If governments are judged according to their ability to prevent epidemic crises, they are equally judged on their ability to avoid expenditures deemed excessive given the existing risk. These political considerations can run counter to the scientific rationale behind the response. This is how, in the case of the 2014 Ebola virus epidemic, the United States became involved: through the declaration by the Liberian President on August 6, 2014, on the threat to national security posed by Ebola, and the danger of the spread of the epidemic to the U.S. soil. In parallel with vaccine research and development, the actions of the U.S., the WHO and the United Nations have focused on treatment of infected patients and the epidemiological securing of burials. In addition, despite a sometimes limited human impact, the economic impact of epidemics involving indirect costs (consequence for certain sectors of activity) has shown a marked increase. However, what characterizes modern epidemics is the duration of the economic « shock », in that it is temporary, as opposed to previous epidemics during which the shock tended to be drawn out in particular due to the persistence of infectious sources. There are many examples of this: among others, the « Spanish influenza » of 1918, the effects of which (company closures, loss of revenue) faded out in 1921, or, more recently, the SARS-CoV epidemic, when the recession, that had been triggered by alerts against travel to Southeast Asian destinations communicated in March, ceased once these alerts were lifted two or three months later. It is of note that during the SARS-CoV epidemic only certain sectors (especially tourism) were affected in Asia and in Ontario, Canada, and not the entire global economy.

Carrying out preventive measures such as staff training makes it possible to limit the number of crises at a lower cost. The cost-effectiveness of such an approach has already been shown. It is henceforth necessary to raise awareness in decision-makers of the importance of prevention relative to risk.

Beyond these constraints, the decision to intervene is complicated by the heterogeneous nature of potential epidemics of the different pathogens. For this reason, and given the absence of technology permitting the prediction of the epidemic potential of a pathogen, prevention which targets the agent is impossible. Prevention must therefore adopt other means, such as training locals in order to improve hygiene conditions. The improvement of transversal knowledge on infectious agents' transmission carries particular importance for the goal of preventing emergence.

Once an emerging agent is introduced into the human population, the pathogen propagation phase within the human population is the key phase in the development of the epidemic, and the one during which it is still possible to act in order to prevent the amplification of the pathogen in the population. The beginning of the propagation phase can be difficult to identify, and improvement in diagnostic techniques as well as the development of rapid diagnostic tests, and even on the field, makes it possible to accelerate detection of an epidemic signal. Moreover, given the key role of certain animal species in the development of new epidemics, the development of animal health surveillance would allow us to further shorten the time lapse between the introduction of the pathogen and its propagation; however, this poses significant problems of wild animal monitoring particularly in Southern Hemisphere countries.

Axis 3: Risk and crisis governance in relation to EIDs

Better knowledge of States' operational modes in dealing with risk makes it possible to better understand their interventions and to better adapt the scientific response. The notion of risk has, little by little, become political, and the State uses risk to govern. Risk management is therefore at the very heart of the State. Recent public health crises have led to risk's taking on a new form, that of the unknown, by the fact of the unpredictability of its appearance and evolution. An interaction between scientific and political approaches to risk is absolutely necessary, in order to better evaluate at-risk situations according to modern methods such as structured decision-making, and to better handle risk-management tools. The changing nature of risks calls for the emergence of a new form of governance, which puts the individual back into the center of the State's action, as well as the confronting of arguments with expert committees. Civil society's participation in risk management should be developed, and observation should once again play its part in crisis response. Younger generations might also take greater part in responding to current crises.

Priority proposals in a nutchell

Seven priority proposals can be outlined as follows:
- Encourage research on the prediction, screening and early detection of new risks of infection
- Develop research and surveillance concerning transmission of pathogens between animals and humans, with their reinforcement in particular in intertropical areas (« hot spots ») thanks to public support
- Pursue aid development and support in these areas of prevention and training for local health personnel, and to foster risk awareness in the population
- Ensure adapted patient care in order to promote adherence to treatment and to epidemic propagation reduction measures
- Develop greater sensitization and training among politicians and healthcare providers, in order to better prepare them to respond to new types of crises
- Modify the logic of governance, drawing from all available modes of communication and incorporating new information-sharing tools

- Develop economic research on the fight against EIDs, taking into account specific determining factors in order to create a balance between preventive and treatment approaches.

## Conflicts of interest

Conflict of interest: none.

## Acknowledgements

We are grateful to the speakers who generously brought their contributions to this seminar: Peter Daszak, Olivier Borraz, Alfredo Pena-Vega, Patrick Lagadec, Maria Van Kerkhove, Allison McGeer, and Benoit Guéry.

The seminar was organized under the Patronage of the Ministries of Social Affairs and Health, Social Services and Environment, Energy and Marine Affairs. This seminar was made possible thanks to the support of the following partner institutions and learned societies:

This seminar was held under the auspices of the French Social Affairs and Health Ministries, as well as that of the Environment, Energy and Marine Affairs. It is organized in the framework of a multiple partnership with ANSES (the French Agency for Food, Environmental and Occupational Health & Safety), the Health Chair of Sciences-Po, the EVDG (Val de Grace School) from the Armed Services Health Division (SSA), the School for Advanced Studies in Public Health (EHESP), the High Council for Public Health (HCSP), the Pasteur Institute of Paris, the French National Research Institute for Sustainable Development (IRD), the Thematic Multi-organisms and Public Health Institutes (ITMOs) - Immunology, Inflammation, Infectiology and Microbiology Institute (I3M) and Public Health Institute (ISP) from the National Alliance for Life Sciences and Health (AVIESAN), the French Microbiology Society (SFM), the Francophone Society of Infectiologists (SPILF), the Paris - Diderot University and the financial support of the BioMérieux Foundation.

We thank Corinne Jadand for helpful organization and management support.